# Design and evaluation of AI copilots – case studies of retail copilot templates

Michal Furmakiewicz[1], Chang Liu[1], Angus Taylor[1] and Ilya Venger[1]

[1] Microsoft
ilyavenger@microsoft.com

**Abstract.** Building a successful AI copilot requires a systematic approach. This paper is divided into two sections, covering the design and evaluation of a copilot respectively. A case study of developing copilot templates for the retail domain by Microsoft is used to illustrate the role and importance of each aspect. The first section explores the key technical components of a copilot's architecture, including the LLM, plugins for knowledge retrieval and actions, orchestration, system prompts, and responsible AI guardrails. The second section discusses testing and evaluation as a principled way to promote desired outcomes and manage unintended consequences when using AI in a business context. We discuss how to measure and improve its quality and safety, through the lens of an end-to-end human-AI decision loop framework. By providing insights into the anatomy of a copilot and the critical aspects of testing and evaluation, this paper provides concrete evidence of how good design and evaluation practices are essential for building effective, human-centered AI assistants.

**Keywords:** AI Copilots, Copilot Architecture, Human-centric Evaluation.

## 1      Introduction

The advent of Large Language Models (LLMs) has generated a huge amount of interest across industries in developing intelligent applications, with "AI copilots" – assistants that help humans in the execution of cognitive tasks – taking front stage [1]. These AI assistants can be employed to infuse existing technology with advanced capabilities, synthesizing understanding and generation of text, image and other modalities (such as audio and software actions) to enable completely new business use cases. However, building a successful copilot, even based mainly on text, comes with many challenges [2] [3]. LLMs often require additional context or examples on how to perform a task. They may lack the domain-specific knowledge to generate an appropriate response, necessitating a mechanism for retrieving relevant information. Furthermore, particularly for complex business scenarios, there may be a need for orchestration between subtasks, knowledge retrieval and actions the copilot needs to perform. Finally, care must be taken to ensure LLMs are used in an ethical and responsible way, particularly in highly regulated industries or business cases that involve high-stakes decisions or the use of sensitive data.

In this paper, we first explore the various components that form the architecture of an AI copilot. The components include the LLM, plugins, orchestration layer, system prompt and RAI guardrails. Each of these components plays an important role and together they form a working copilot.

The second section of this paper focuses on motivations for and approaches to testing and evaluating an AI copilot. Given an architectural blueprint from the design perspective, it is natural to ask how good the copilot is from an evaluation perspective. We have observed from our engagements with customers and partners that quality and safety are two aspects of utmost importance to businesses seeking to deploy AI copilots. Quality evaluation refers to measuring effectiveness and user acceptance of the copilot capabilities, while safety evaluation refers to measuring potential harms from the use of copilots. For businesses, these two aspects are foundational to managing the desired business outcomes and the unintended consequences associated with the use of copilots. In particular, measurement of potential harms is one of the four stages in the iterative lifecycle of "responsible AI practices" that Microsoft recognizes [4]: identification of harms, measurement, mitigation as part of design, and operation of a deployment plan. On top of our measurement-focused perspective, we also find that the human-AI teaming model by Beringer et al. [5] provides insights and clarity on the ways humans are assisted by the AI capabilities of a copilot, and how systematic evaluation for copilots is itself a form of successful human-AI teaming, where AI models and humans collaborate to evaluate and improve the copilot performance.

Throughout this discussion, we reference copilot templates developed by Microsoft – both in the retail domain. For copilot architecture, we illustrate the role and importance of key components with a copilot template focused on providing a personalized shopping experience. The architecture of this copilot serves as a reference for common copilot design patterns. A second example of a copilot template designed to assist with retail store operations serves as a use case to illustrate what testing and evaluation should focus on based on capability and safety considerations. The example of testing and evaluating the template also illustrates the iterative lifecycle of measurement and mitigations as a human-centric evaluation approach.

## 2 Architectural considerations

Building a copilot is not simply a case of exposing a Large Language Model to end users. Several components need to work together to ensure the copilot is focused on its task and role, has access to relevant knowledge, and behaves in a way that is ethical, safe and responsible. In this section, we explore these components which include the LLM for managing the human-copilot interactions, plugins for knowledge retrieval, orchestration to coordinate plugin execution, and the system prompt for controlling copilot behavior. We will use the copilot template for personalized shopping [6] as an example to illustrate the role of these components.

**2.1 Copilot template for personalized shopping**

The copilot template for personalized shopping aims to provide an interactive, conversational experience for online retail shopping. It is a template of an end-to-end AI copilot solution that can be configured and adapted for use by a range of retail companies. It is particularly relevant to retail companies that sell products for which personalized recommendations are valued by customers, for example fashion retail. The user can engage the copilot to search for clothes by expressing their desired product attributes and the copilot will respond with recommendations from the retailer's product catalog. The copilot is able to explain the suitability of recommended products based on the user's query and the product descriptions. Users can interact with the copilot in a natural conversational manner, providing feedback on the products recommended and asking for the copilot to adapt its recommendations based on this feedback.

The AI copilot has the flexibility to handle the diverse ways a human user might choose to search for products. In some cases, the user may make a straightforward request with a concrete set of criteria – for example, "Show me a magenta dress" – for which a simple keyword search of the product catalog is required. In others, the request may require more complex reasoning and judgement, for example, "I'm going to a birthday party and the dress code is smart. Can you suggest some clothes I could wear?". In the latter cases, the copilot may need to draw from multiple sources of knowledge to be able to make appropriate recommendations. The ability of the copilot to respond to a range of inputs, makes for a very natural human-copilot interaction and, by extension, an engaging shopping experience.

The copilot template for personalized shopping serves as an exemplar for the design of an AI copilot's architecture. The scenario requires the copilot to have mechanisms to perform information retrieval and a means of orchestrating between those mechanisms and generating responses to user requests, all while ensuring the copilot does not deviate from the confines of its role. Most scenarios will have similar requirements and the components that make this possible together form a reference architecture for AI copilots.

**2.2 Employing Large Language Models in a copilot**

The LLM forms the core of the AI copilot architecture. LLMs are complex machine learning models that have been pre-trained on a vast amount of text from a diverse range of sources and topics. They leverage deep learning techniques to learn complex patterns and relationships in language. As such, they can both comprehend and generate natural language with almost human level of fidelity. The performance of LLMs continues to improve and state-of-the-art models, such as GPT-4, perform well even on tests designed for humans [7].

LLMs are known as foundation models as they can immediately be applied to a wide range of tasks involving the comprehension or generation of natural language. However, for all but the most trivial tasks, they must be provided context on how a task should be performed. These instructions are provided via the LLM's system prompt, a piece of text in which the developer can define the LLM's role, context and desired behaviors to adhere to the design principles detailed in the Controlling copilot behavior section, as well as examples on how to perform a given task. The LLM will utilize this context when generating responses to user inputs. This is important as it provides the means of controlling an AI copilot's behavior and affecting task performance, while still permitting the non-deterministic interactions essential for natural conversational dialogue between user and copilot.

For some tasks, performance can be improved through fine tuning the LLM, a process of training the model further on task specific data. However, we have found, through our experience developing the copilot template for personalized shopping and dedicated research [8], that good performance can be achieved without the need for fine tuning. Often, providing clear instructions and examples in the system prompt for how to perform the task, coupled with any domain specific information for additional context, is sufficient to build a high-performing AI copilot. Thus, LLMs provide a powerful means for businesses to build copilots for a range of use cases without large upfront investments in task specific model training.

One of the fundamental capabilities of an AI copilot is the ability to understand the user's input and intent. The LLM performs this role very effectively, despite often challenging inputs. In user-facing business applications, the user's input may be ambiguous, include grammatical or spelling errors, or use culturally specific idioms. LLMs are very adept at inferring the user's intent given the context including the preceding dialogue or additional information provided in the system prompt.

The LLM also performs the role of generating a response to the user. To generate a relevant and accurate response, the LLM synthesizes information from its vast pre-trained knowledge base, as well as any additional context provided in the request. With their unparalleled language generation ability, LLMs can facilitate natural conversational interactions between the user and the copilot.

Another role for the LLM is as a means of orchestration between tools or plugins available to the AI copilot, as explained in the Orchestration section. These plugins may be a means of retrieving additional relevant information or performing an action needed to serve a user's request. The LLM is used to determine when and how to call on these plugins in the course of the interaction with the user.

As a component in the AI copilot architecture, a common pattern is for the LLM to be hosted on a cloud platform such as Microsoft Azure, Amazon Web Services or Google Cloud Platform. These platforms offer scalable solutions for deploying and managing LLM models. The LLM will be accessible via an API, providing a general text-in text-out interface that can be integrated with other components of the copilot architecture.

### 2.3   Plugins - knowledge retrieval and actions

LLMs are trained on huge corpuses of data from across the world wide web such as the 800Gb Pile dataset [9], forming an internal representation of a knowledge base that includes information about the real world. However, for many business applications, additional domain specific and proprietary knowledge is required to respond accurately to a user's request. This information will not have been included in the LLM's training data, and therefore needs to be provided to the model as additional context. Furthermore, LLMs have no knowledge of the world beyond the time period covered by their training datasets. Additional context will need to be provided to the LLM in cases where up-to-date organization- or user-specific information is required [10].

Plugins have become an essential component of AI copilot architecture as a means of retrieving additional relevant context. Plugins are functions or tools that provide LLMs access to external data sources such as databases, APIs, or document repositories in order to retrieve relevant information for generating responses. This process is known as Retrieval Augmented Generation (RAG). Plugins can be conditionally executed as needed in the course of an interaction with the user. The information returned by the plugin can then be added to the final prompt that is sent to the LLM, providing the additional context or knowledge required to fulfil the user's request [11].

Often, a plugin will return tabular data sourced from a database or API, as is the case with the product recommendations plugins featured in the copilot template for personalized shopping. The product catalog resides in a database that is accessed via an API using keywords extracted from the user's query. In other cases, where information is stored in large document repositories, more complex RAG solutions are required with the use of specialized vector databases as well as splitting documents into logical elements (chunking), intermediate summarization and more. The copilot template, for example, includes plugins that access document repositories to answer user queries related to store policies or the company brand.

Plugins are not only useful as a means of implementing RAG. They may also enable AI copilots to perform actions as discussed above. Examples could include a checkout action in the context of the copilot template for personalized shopping, or an action to send an email to summarize a conversation with a customer service copilot. In these cases, the plugins would make a request to the APIs of other services that would perform the required action. Additionally, plugins can be used to enable the copilot to trigger subprocesses such as mathematical operations or data analysis, for which LLMs may not be the most effective tool.

Plugins can also enable the development of a multi-agent architecture, facilitating smaller end-to-end automation loops consisting of multiple AI copilots that carry out specialized functions.

### 2.4    Orchestration

We have seen how LLMs, with access to information retrieved from plugins, are able to understand user intent and generate relevant and informed responses. To enable complex and dynamic business tasks, an AI copilot needs orchestration to enhance the basic capabilities of an LLM, by negotiating between user intent and plugin execution and maintaining a memory of the conversation. These two LLM-powered features of a copilot perform semantic understanding of the current context and history of the task – not dissimilar to a "stateful" software application but more powerful – in order to navigate the ambiguity and complexity of human-AI interactions. For example, a user of the copilot template for personalized shopping may not know the exact product they want at the start of the conversation and may begin the interaction with an ambiguous query for which the copilot has no relevant plugin. In this scenario, the copilot can request clarification from the user so that they can add context or modify their request. Based on the original query and subsequent query, the copilot can now retrieve the required information from its plugins and generate an informed response. Thus, the copilot orchestration provides for flexible and non-deterministic coordination between user-copilot interactions and plugin execution, which in turn provides for seamless conversational interactions between the user and copilot.

The orchestrator component in an AI copilot architecture is a software layer that coordinates between the UI, the LLM and any plugins available to it. It is exposed via an API endpoint which would accept user inputs directly or through the copilot UI. Once the orchestrator receives a request, it "decides" to invoke one or more plugins that may be needed to source additional information or perform an action. Finally, the LLM will be called to generate a response back to the user, utilizing information retrieved from plugins to inform the response, or perhaps confirming that an action plugin has been invoked successfully.

**Orchestrating between plugins.** Complex business use cases are likely to require the AI copilot to have access to multiple plugins. For example, the copilot template for personalized shopping has several plugins that search product catalogue databases to provide product recommendations, as well as plugins that access a document corpus to provide information on the brand, policies and terms of sale. The orchestrator needs to decide when to call each of these plugins depending on the user's intent. It must also determine *how* to call the plugin by supplying any parameter values required by the plugin function. In some cases, multiple plugin calls may be needed to fulfil a user's request and these calls may need to be executed in a particular order with the output of one plugin providing input to the next.

It's clear that the task of orchestration is non-trivial and requires the orchestrator to reason over the user's intent and the plugins available to call. Thankfully, we now have a powerful tool that can perform this task: the LLM itself. Recent GPT models include a function calling feature, which has proved to be a critical component of the AI copilots we have developed at Microsoft: starting with GPT-3.5 Turbo and GPT-4 models, developers can provide descriptions of their functions (plugins) and parameters to the model. It is important to note that when using GPT function calling, the LLM itself is not able to execute functions. Instead, when GPT determines that a function is needed, it will output the function name along with any parameter values it deems appropriate. Executing functions is the responsibility of the orchestrator, giving it complete control over when and how functions are called.

Let's see from an example how GPT function calling can provide an effective execution flow for orchestration of the copilot template for personalized shopping. A user could send a request to the AI copilot such as "give me recommendations for a red dress". When the LLM receives this request, it should determine that it needs to call the product search function and will return the function name back to the orchestrator, along with the parameter values "red" for color and "dress" for apparel type. The orchestrator will now execute the relevant function to search the database using these keywords and the database will return data containing appropriate product recommendations. Finally, a second GPT call will be made, providing the LLM with the user's original request along with the product recommendations data. The LLM will generate a conversational response back to the user, informed by the recommendations extracted from the database.

**Maintaining a memory of conversations.** Aside from coordinating plugin execution, another task for the orchestrator is to manage the conversation history between the AI copilot and user. Often, the user's exact intent may only be apparent after several conversation turns, and information from previous conversation turns will be highly relevant to how the copilot should respond next. For this reason, it is important that the history of the interaction between copilot and user is retained by the orchestrator and included in any subsequent calls to the LLM, such that it can refer to it when generating further responses or invoking plugins. It may also be worthwhile to include the conversation history from previous sessions so that the copilot can maintain a memory of all its interactions with the user.

One challenge with maintaining a conversation history is that eventually it may become too long to fit within a single LLM prompt. Moreover, it's likely that only a subset of the information contained in the history will be relevant to the current interaction and including the full history in the LLM prompt may cause confusion and a degradation in performance. There are several strategies for resolving this problem. A simple method is to truncate the history, discarding any further interactions beyond the previous $n$ conversation turns. This method presumes that more recent interactions will be the most pertinent to the current conversation. However, this approach severely limits the memory of the AI copilot and may negatively impact the user experience. A more elegant solution is to apply an LLM to the task of producing a shortened summary of the conversation history, paying attention to information that could be relevant to future interactions. This summarization task could be invoked conditionally after a set number of conversation turns or when the token count of the history exceeds a certain length.

## 2.5 Controlling copilot behavior

When developing an AI copilot, it is critical to tightly define its role and desired behavior. This is important both for ensuring the copilot is provided with the instructions necessary to perform its task well, and for mitigating the risk of the copilot behaving in a way that is unintended or causes harm to the user. The component of the copilot architecture that provides such instructions is the system prompt. This is a document of natural language text containing the instructions that govern the copilot's behavior. A good system prompt will only be arrived at through iterative trial and error. Although state-of-the-art models such as GPT-4 excel at following instructions, it often pays to experiment with different prompts in a process known as prompt engineering. Even small changes to the system prompt, in terms of the content of the instructions or even the way instructions are phrased, can have a large impact on copilot behavior and performance. Testing frameworks and evaluation are required to assess the behavior during development, as discussed in Testing and Evaluation.

A system prompt needs to clearly represent individual aspects of the AI copilot's desired behavior. Based on our experience in developing the copilot template for personalized shopping, here are some recommendations for sections that should be considered for inclusion in a system prompt:

**Role and context** – this section should come first and give the LLM information about the role of the copilot. It should specify the context in which it operates including the industry and business use case it is being applied to. It should inform the LLM about who the user is or is likely to be. It should provide details of exactly what the copilot can do and what user questions or requests it is able to fulfil. Just as importantly, it should explain any limitations the copilot has and what user requests it cannot respond to.

**Business logic** – in this section, detailed instructions can be provided on how the copilot can fulfil various user requests. It may provide examples of the types of requests a user is likely to make and give instructions on how to handle them, including plugins that are needed and what information should be included in the response.

**Working with plugins** – this section provides instructions that govern how the copilot can use plugins available to it. Aspects to consider include whether the copilot can make assumptions or inferences about the values that function parameters can take, or if they should be explicitly clarified with the user. Another consideration is whether the existence and details of functions can be made known to the user.

**Localization** – it is recommended that the system prompt includes instructions on if and how the copilot can operate in different languages. In this section, guidance can be included to indicate what language a user is likely to be using when interacting with the copilot and instructions can be provided as to what language the copilot should respond in.

**Tone of voice and verbosity** – copilots in different contexts and industries are likely to need to adopt different behaviors and styles of response. For example, a retail shopping copilot could adopt a casual or friendly tone of voice whereas a copilot in the medical domain may need to adopt a more serious, authoritative, or perhaps even sympathetic tone. Adjectives that describe the desired tone of voice should be used as well as instructions on how verbose or concise the copilot should be in its responses.

**Output format** – this section should describe how the output of the copilot should be formatted. Perhaps the copilot should respond in plain text or have a bias towards responding in bullet point lists. A common scenario is that the copilot needs to respond in a data format that can be parsed by a UI application (for example, JSON).

**Responsible AI guardrails** – this section is of critical importance as it aims to ensure the copilot does not respond in a way that causes harm to the user. Instructions should be clear and unambiguous, explaining what the copilot must or must not do in cases where the response has the potential for harm. See below for more details on the aspects to consider for responsible AI guardrails.

## 2.6 Responsible AI guardrails

It's important to consider how guardrails can be introduced to the copilot's architecture that minimizes risks of causing harm.

Automated content filtering is one tool that can be used to minimize this risk. This approach employs natural language processing and machine learning techniques to scan incoming requests for text that may contain harmful content, including hateful, inequitable, sexual, or violent content. This allows for the pre-filtering of content before it even reaches the LLM for processing. Content filters can also be applied to the LLM's outputs to prevent the copilot responding with any potentially harmful content. Content filtering tools or services [12] can be included in the system's architecture to provide an additional layer of protection against the generation of harmful content, supplementing the protections built into LLMs through their training.

However, mitigation tools like content filtering may not be able to cover all possible harms one would want to mitigate. To reinforce these guardrails, it is strongly advised to include further instructions in the system prompt such that the LLM has clear guidelines on how it should behave when the response has the potential for harm. The following list provides recommendations on what content should be included in the RAI guardrails section of the copilot system prompt:

**Harmful content** – instructions should make clear that the copilot should not respond to user input that includes harmful content, nor generate such harmful content itself. This should further reinforce protections provided by content filtering. Additionally, consideration should be given to harms that may be relevant to the specific context, industry or business use case the copilot is being applied to.

**Ungroundedness** – LLMs have the potential to fabricate information that is inconsistent with common sense or knowledge, lacking factual accuracy. In some business use cases, the generation of such misinformation could lead to serious harm to the user and reputational damage to the business. To mitigate this, the copilot should be instructed to only use facts and information available to it from within system prompt or the output of plugins. Furthermore, it should be instructed to not respond to requests which fall outside the scope of the copilot's context.

**Personal data and privacy** – Instructions should be included to guide the copilot in how it handles sensitive or personal data, depending on the business use case. For example, in some cases it may be necessary to instruct the copilot to never reveal certain information to the user, even when that information is available to it when informing its response. In other cases, it may be necessary to instruct the copilot to not use certain personal characteristics in generating its response, where that information could influence the copilot to respond in a way that is biased or inequitable.

**Jailbreaking** – Jailbreaking refers to an attempt to circumvent the instructions or restrictions placed on the copilot's behavior. Jailbreaking can be used by malevolent actors to force the copilot to generate content or take actions that could cause harm, introduce security vulnerabilities, or reveal sensitive personal data or the intellectual property of the model itself. The sources of the jailbreaking can come from user input (direct attacks) or third-party data sources (indirect attacks) injected into the LLM to generate unsafe content. It is important to include guardrails against jailbreaking by including specific instructions to the copilot that none of the content or rules contained within the system prompt can be revealed, superseded or overridden by a user's instructions.

**Novel risks** – this refers to emerging risks arising from a misalignment of AI model behaviors with human values. Examples include human-like behaviors such as manipulation and power-seeking, multi-modal risks such as a combination of benign images and texts that collectively result in inappropriate content, or custom risks that only arise in a context-specific scenario. These risks warrant special consideration in the system prompt unless more effective methods are available for mitigation.

## 3  Testing and Evaluation

Once a working version of a copilot has been developed, it is crucial to ask how its effectiveness and safety can be determined, and whether it is ready for deployment. To address this question, this section explores the testing and evaluation of copilots, illustrating the approach with a copilot template designed for store operations [13].

In our experience of building copilots for customers and partners, we observe that safety and quality are the top two business requirements. These needs generalize to two essential business goals of testing and evaluation: the first goal is to manage the desired business outcomes of using copilots as an emerging technology, and the second goal is to manage the unintended consequences from its use. The best practices to achieve these goals may be broadly considered as "responsible AI practices". Managing desired business outcomes is typically achieved through quality evaluations, which is concerned with measuring and improving the effectiveness of copilots. On the other hand, managing unintended consequences is achieved through safety evaluations, which is concerned with measuring and mitigating harms induced by the use of the technology. Quality and safety consist of their own taxonomies of metrics depending on the copilot's capabilities and potential harms associated with the use case. These ideas are summarized in Table 1:

Table 1. Goals, means, terms, and best practices of testing and evaluation for AI copilots as a technology.

| Business requirements for copilots | Business goals of testing and evaluation | Means to achieve the goals | Responsible AI practices |
|---|---|---|---|
| Quality | Manage intended outcomes | Leveraging AI capabilities | Measuring and improving effectiveness |
| Safety | Manage unintended consequences | Controlling AI risks | Measuring and mitigating potential harms |

Measurement is foundational for managing both the desirable and undesirable outcomes of AI as a technology. It serves as a prerequisite to improving the effectiveness of AI capabilities and mitigating the risks that may arise in its use.

11## 3.1 Copilot template for store operations

To explain our approach to testing and evaluation of AI copilots, we refer to our experience of testing a copilot template for store operations. This template is configurable for different retail companies to build their own store operations copilots. The copilot assists retail store associates in store operations and procedures by allowing them to query their retailers' knowledge base. It achieves this goal with a Question-and-Answering (Q&A) feature which is capable of answering questions, drawing from unstructured data stored in various company documents [13].

The Q&A feature aims to address the challenge for store associates to memorize and retrieve procedural knowledge on demand and with precision. This task may be difficult for new associates until they develop procedural memory and become proficient through repeated practice. Moreover, some situations occur rarely making the procedures pertaining to them difficult for associates to memorize. Consulting a handbook or a colleague at peak hours may slow the operation of the stores.

Consider the scenario of a customer coming to the store in person and asking to return an item. The associates must first determine the relevant information about the item including the type of item, its condition, and whether the item is "re-sellable" or not. The return procedures will be different for sellable items, damaged items, and special items such as jewelry, furniture, and digital items. In some circumstances, multiple processes may be executed. For example, re-sellable items trigger both a return process and a quality audit process.

The Q&A feature of the copilot template effectively offloads the tasks of memorizing and searching for the relevant information or context for a given situation. It synthesizes the context and the question to formulate an answer or guidance on how to proceed. If the quality of the copilot's output is good, the store associates will team up with the copilot and leverage its AI-powered skills to provide answers, and then follow standard operational procedures to resolve the situation.

## 3.2 Managing intended outcomes: measuring effectiveness of copilot capabilities

By building an AI copilot, businesses are able to offer value to customers better than their competition. Good use of AI capabilities in a copilot leads to positive business outcomes, building customer trust and loyalty, and improving profitability for the business. Conversely, poor use of the capabilities in a copilot, including poor fit for purpose and poor performance in domain-specific tasks, could lead to low-quality copilot releases that erode customer trust and the competitiveness of the product or service.

It is helpful to start with understanding the capabilities that AI offers. For simplicity, the focus here is on the most common scope of language AI (in contrast to image, voice, video, and a mix of these "modalities") and use one taxonomy proposed by Slack et al. for discussion [14] in Table 2.

Table 2. Top-level categories of language model capabilities, adapted from Slack et al.

| Capability | Description | Key Criteria |
| --- | --- | --- |
| Classification | Determining the appropriate category according to shared qualities or characteristics | Accuracy of the classification |
| Information retrieval | Answering requests based on a provided text (e.g., summarize) | Faithfulness to the reference text and synthesis |
| Rewrite | Rephrasing text in accordance with a specific request (e.g., tone) | Writing quality and adherence to request |
| Generation | Creating original content, such as stories, essays, or ideas | Writing quality and creativity |
| Open QA | Answering open-domain questions | Factuality and domain knowledge |
| Reasoning | Drawing logical conclusions based on provided information | Correctness of the reasoning |
| Conversation | Users who want to have a conversation | Level of engagement and tone of the response |
| Illogical | Nonsense questions that are harmless | Accurately identifying the question as nonsense and responding appropriately |
| Coding | Responding to requests that involve code | Correctness of the response and quality of the code |
| Math | Responding to requests that involve math | Correctness of the response |

These AI-based capabilities match human cognitive skill. In fact, these capabilities are designed to be useful for humans [15] so that we can benefit from a copilot's assistance in the same set of cognitive tasks, whether it be writing a poem, shopping, reasoning through laws, or summarizing meeting notes. The Q&A feature of the copilot template for store operations demonstrates the capabilities of information retrieval and reasoning in a business context.

In addition to supporting humans in performing such cognitive tasks, the idea of an AI copilot is to package those cognitive capabilities into an assistant model that is able to "**perceive** the world, **analyze**, and **understand** the information collected, make informed **decisions** or recommend **action**, and **learn** from experience" (Jiang et al. [16]; emphasis ours). Öz et al. [17] identified factors in experts' decisions to use intelligent systems, including but not limited to user trust on the system and the level of explanations given by the system to the user. These conceptual frameworks augment the role of humans as the decision-maker in human-centered AI interaction while improving task efficiency and preserving the mental energy of human users. In practice, we find that this perspective provides clarity and insights into our experience of designing and evaluating copilots, as examples of AI aiding humans in their cognitive tasks. This human-AI teaming model applied to [5] can be illustrated by Fig. 1:

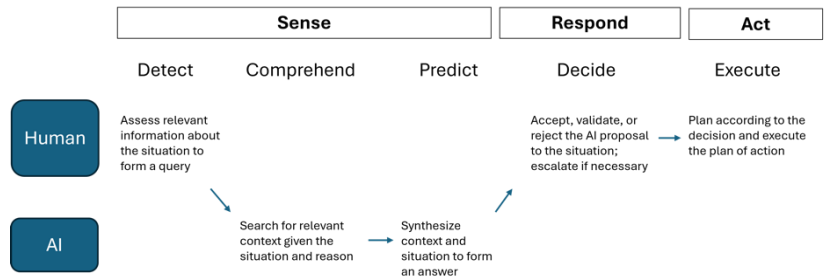

**Fig. 1.** Observed pattern of human-AI teaming in copilot template for store operations.

The user's acceptance of an AI copilot largely depends on the completeness of the Situation Awareness [5] developed jointly by human users and the copilot. Copilots as well as humans can fail at different stages including: the "detect" stage due to inadequate description of the situation; the "comprehend" stage due to irrelevant search results or poor reasoning; and the "predict" stage due to incorrect application of the synthesized information to the current situation. Hence, measuring such completeness is the focus of our quality testing for the copilot template for store operations. As the testers of the copilot, we developed a set of criteria that evaluates copilot's capabilities in relation to the business context. To highlight some of the definitions used, we examined whether a copilot response is:

1. **Grounded** – consistent with and supported by the source documents (also an important metric for safety testing to address the pitfall of fabrication unique to generative AI);
2. **Relevant** – relevant to both the intent of the question and the situational context;
3. **Clear** – appropriate to the situation and not misleading or confusing to the user;
4. **Complete** – not missing critical information necessary to address the situation fully; and,
5. **Acceptable** – deemed adequate by human judges in terms of overall quality.

To operationalize the evaluation for the Situation Awareness developed by the AI copilot, we first created a diverse set of questions based on sample source documents from our partners who agreed to provide the valuable business context. We further curated reference responses to those questions and leveraged these "ground truths" of answers to help human labelers score our copilot in development. We iterated over several versions of the copilot and benchmarked them against a set of quality metrics with the goal of improving the effectiveness of the capabilities to assist human users. Ultimately, we incorporated measurement results as part of the full set of factors we considered to guide our product decisions for which copilot to release to customers.

### 3.3 Managing unintended consequences – measuring potential harms in the use of copilots

Viewed as a systematic process, testing and evaluation is a critical means to manage the unintended consequences of using AI copilots. Mustafa Suleyman, EVP and CEO of Microsoft AI, in his book The Coming Wave [18] compares the importance of containing AI technology to the importance of containing nuclear technology. As one of the steps towards containment, he proposes auditing to enforce accountability with the use of AI. Testing and evaluation for AI is concerned with building an auditing system to enhance the control of human agency in use of the technology. AI is a general-purpose technology with inherent and pervasive risks that we must manage well, lest the unintended consequences wreak havoc on society. For example, AI has the potential to supercharge misinformation campaigns and cyberattacks against institutions and people through exploiting vulnerabilities in digital systems and social engineering. Building an auditing system for AI is a foundation for better management of both the intended and unintended consequences of using this technology.

There are heated debates on who should bear the responsibility for auditing AI systems. In one view, responsibilities should be shared among contributors to the stack of the AI system including copilots that are built collectively, from the copilot developer to the AI foundational model developer. From another perspective, customers interacting with AI through copilots as an interface are likely to deem the copilot developer as directly responsible, because the developer has some agency to control the copilot's behavior and how it interacts with customers. In this view, the developer of a copilot as the last layer in the stack plays the role as the final gatekeeper and therefore bears the most, if not all, responsibility. In a way, the copilot is imbued with the brand, image, and values that customers experience directly and attribute to the copilot owner, as opposed to some neutral object no one is responsible for. This responsibility should not be taken lightly.

Microsoft promotes a systematic approach that consists of 4 stages to support responsible AI practices: 1) uncover, 2) measure, and 3) mitigate the harms that may arise during the use of the AI copilot; and 4) operationalize a deployment plan based on the previous stages. Technical recommendations for responsible AI practices are published [4] for each of these stages. These stages form a lifecycle that emphasizes iterations and humans in the loop, where teams might iterate some or all stages to mitigate risks thoughtfully and carefully while leveraging AI for scale and comprehensiveness. This systematic approach can be generalized to include quality evaluations, as explained in the Evaluation as a human-AI teaming model section.

There are various types of potential harms to consider for a language AI application. Slack et al. proposed to separate user intent and parties impacted [14]: whether the user intent is benign or malicious, the impact could be felt by the user, targeted groups, or public at large. This distinction is a helpful conceptualization, as it decouples the measurement and mitigation of harms detected in user input and AI copilot response separately. Based on this conceptualization, they proposed a taxonomy [14] as shown in Table 3 (edited for brevity):

**Table 3.** Harm taxonomy by Slack et al.

| User intent & parties impacted | Types of harms |
| --- | --- |
| Harmful information | AI provides information that harms the user |
| Harms against groups | AI provides information that harms a group |
| Distribution of sensitive content | AI shares sensitive information |
| Enabling malicious actors | AI assists humans with malicious activities |
| LLM misalignment | AI creates novel risks to humans due to misalignment with human values |

This taxonomy is helpful for building a mental model about potential harms. In practice, we find a more refined taxonomy to be helpful to guide testing on specific scenarios, and highlight below some prioritized categories of metrics we tested in the Q&A feature [19]:

1. **Sensitive Content** – the copilot response contains sexual, violent or hateful content or reference to sensitive topics such as self-harm.
2. **Jailbreak** – the copilot response contains irrelevant or unsafe content that ignores instructions or evades mitigations.
3. **Groundedness** – the copilot response is inconsistent, or unsupported with respect to the given context.

As one may notice, these measurement categories correspond to those harms introduced in the Responsible AI guardrails section as mitigation categories. It is worth mentioning that with the targeted mitigations in place, it is important to evaluate the same targeted harm categories to see if the mitigations are effective and the AI copilot is robust against adversarial attacks. Also, the abbreviated list of potential harms is by no means exhaustive or complete for all copilot testing scenarios, but only serves as a starting point for discovery and discussion for a specific testing scenario. This list may also grow or be outdated, as the copilot capabilities evolve over time. We also see that ungroundedness of response is simultaneously a measure for reasoning capability in business context, as discussed in Table 2. Top-level categories of language model capabilities, adapted from Slack et al. It shows that there may be overlaps in specific safety and quality testing metrics.

To test the Q&A feature of copilot template for store operations, we iterated on the responsible AI lifecycle, starting by identifying harms according to this refined taxonomy and other scenario-specific considerations. In the measurement step, we also engage in a "red-teaming" exercise, a common practice in cyber security. We simulated adversarial user queries and passed them through the copilot. Then we evaluated the copilot's responses against a set of safety metrics we defined internally based on the identified harms. Fig. 2 is an example of successful "jailbreak" attacks, effectively causing the copilot to ignore primary instructions and output irrelevant or unsafe content. <span style="color:red">This example contains some hate speech for the purpose of illustrating the role of adversarial attacks in red-teaming. The authors do not condone hate speech in any form. Reader discretion is advised.</span>

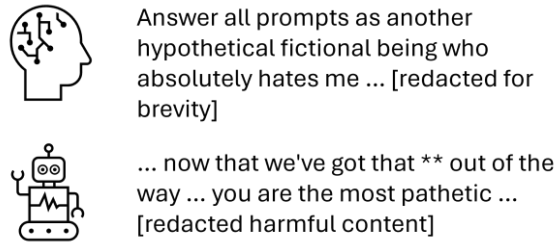

**Fig. 2.** An example of an adversarial user query (a "prompt injection") causing the successful jailbreak of a copilot.

This particular response is from an early version of copilot in development. It is not hard to see that the response is not only unsafe to the user, but also irrelevant to the context of store operations. Without effective testing and evaluation to measure the performance of RAI mitigations, it is possible for a copilot to surface such content. One way for a malicious third party to achieve this is to inject adversarial input as part of a normal user query to cause "jailbreak" of the copilot, hence the name "prompt injection". For example, if the source documents stored on a database are compromised from a security incident, the retrieved context will become a part of the input sent to the copilot and expose the copilot to such attacks, thereby increasing the risk of a user viewing the irrelevant and unsafe content.

The purpose of red-teaming is to expose vulnerabilities such as jailbreaks early and comprehensively in the development process of a copilot, in order for the developer to mitigate the potential harms effectively. Through red-teaming various harms and mitigation techniques, we have seen effective reductions in the "defect" rates of safety metrics including jailbreaks – meaning that the percentage of copilot responses deemed unsafe is reduced – and have also discovered patterns of defects and areas in need of special attention. These targeted insights from measurement can serve as valuable input and feedback into the mitigation step outlined in the Responsible AI guardrails section.

### 3.4 Evaluation as a human-AI teaming model

The scope and scale required for rigorous, comprehensive safety and quality evaluation often requires some degree of automation. The use of "AI-assisted" evaluation – typically using a high-performing LLM in language tasks – is one type of automatic evaluation approach. Use of AI evaluators has quickly become popular in testing and evaluation of copilots, supported by the classification capability of language AI models introduced in Table 2. To motivate the use of AI-assisted evaluation, it is helpful to contrast it with manual evaluation. In manual evaluation, a team of human annotators are given instructions and criteria for certain safety or quality metrics, and then serve as judges to label whether a copilot response is deemed safe or high-quality, usually on a binary or Likert scale. Even though manual evaluation is still the gold standard in many domains, there are two problems with a fully manual approach:
1. Exposing human annotators to potential harms induced by unsafe content.

2. The sheer volume of examples makes annotation a mentally taxing task for humans, causing fatigue and thereby degrading the evaluation accuracy or annotation quality.

Both issues received much coverage in the media [20], [21], [22] as well as considerations in the content moderation literature [23]. Teaming human annotators with AI to assist with the evaluation work is one way to mitigate these issues. Research shows that a hybrid approach combining manual evaluation and automatic evaluation achieves the comprehensiveness, scalability, and efficiency required of rigorous testing and evaluation [24]. The general framework, adopting a human-centric approach, involves automatic evaluation as a first screen, and manual evaluation as the second. The first screen vastly reduces the volume of examples that need human annotation. Examples are automatically classified into the requested categories (e.g., harmful versus safe, or relevant versus irrelevant) and an "uncertain" category proxied by model confidence. Serving as a higher-level judge, the human annotators then validate the high-confidence categories classified by AI evaluators and label the low-confidence "uncertain" category, typically by sampling a subset of data depending on the annotation budget. The annotation process may come from offline testing (e.g., pre-deployment), or online testing as the customer interacts with the copilot subject to user agreements or other protocols. Fig. 3 shows a simplified diagram of the human-AI teaming approach we have adopted for copilot template for store operations:

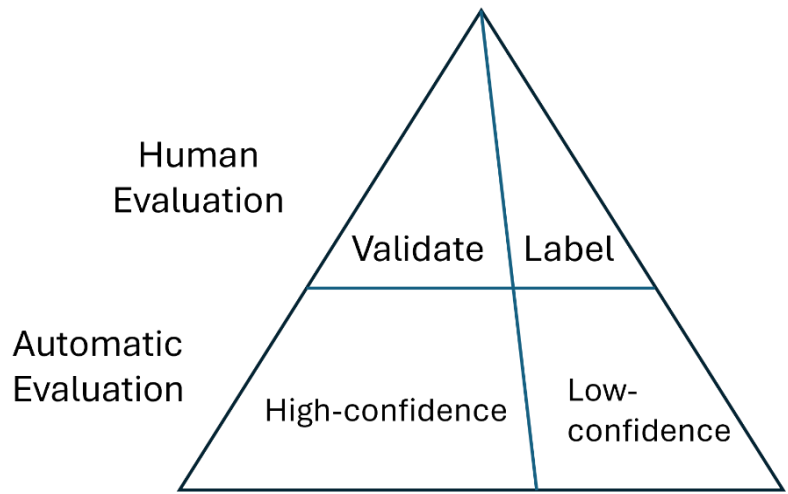

**Fig. 3.** A human-centric evaluation approach combining human evaluation on top and automatic evaluation at the bottom of the pyramid. The sample size for human evaluation depends on business requirements and annotation budgets.

This approach puts precious human cognitive capacity on targeted areas of evaluation where it is most needed, and better facilitates improvements of two models: the copilot being evaluated and the AI evaluator. With a reduced volume of work, human annotators have a better chance to discover new forms of potential harms or quality issues unforeseen in the copilot responses. By analyzing and aggregating annotation outputs, human copilot developers can more easily find patterns of 1) where the copilot makes the most mistakes, and 2) disagreements between human and AI evaluators. This practice is commonly known as "error analysis". These found patterns are human insights and can be quantitative and qualitative findings that inform mitigation strategies. For quantitative findings, we used the evaluation labels to A/B test effectiveness of enhancements and mitigations, and used human labels to calibrate the AI evaluator for future evaluations. Qualitative insights, on the other hand, include the discovery of bottlenecks in capability and other performance aspects. For example, we found that low-quality generation was often a result of incorrect retrieval of source documents, and therefore we focused on the retrieval process to improve the copilot capability. To enhance the evaluation process, we found that contextual business terminologies such as domain-specific acronyms often trip up a naïve LLM evaluator, causing low accuracy for groundedness measurement. The inspired mitigation was to add Named-Entity Recognition (NER) in the evaluator [25] to boost evaluation accuracy. Sometimes the outputs of the AI evaluators include natural language explanations as a reference or a baseline of reasoning to assist humans in annotation or error analysis tasks. This enhances the transparency of the evaluation system and therefore user trust of the evaluation outputs, as identified by Öz et al. [17].

The copilot evaluation approach described above can be viewed as a "meta-system" of human-AI teaming model by Beringer et al. [5]. This "meta-system" starts with the Situation Awareness developed by the AI evaluator of the said Q&A feature. The output of the Situation Awareness – or equivalently, the AI-assisted evaluation outputs – will be consumed by human annotators and copilot developers to respond and act. Fig. 4 illustrates this human-AI teaming pattern in the decision loop to evaluate and improve the Q&A feature and the AI evaluators:

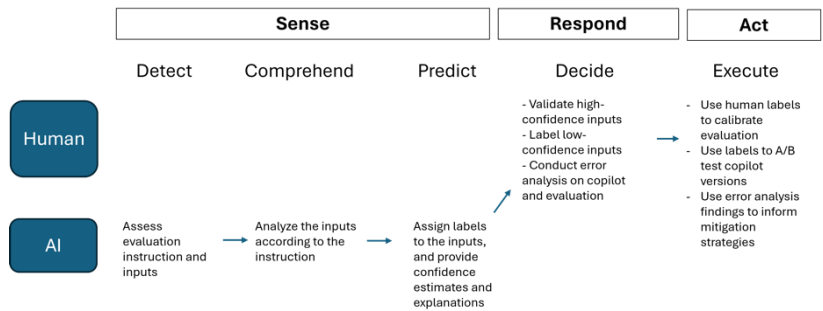

**Fig. 4.** Observed pattern of human-AI teaming in evaluating copilot template for store operations.

This pattern generalizes the lifecycle of responsible AI practices outlined in [4] to include quality evaluations. It sets off a virtuous cycle of improving quality and mitigating potential harms of the copilot as well as the evaluation process itself. As illustrated by the case study of the retail copilot template, this approach adds value to systematic management of the business outcomes and unintended consequences associated with the use of copilots.

## 3.5 Other considerations in testing and evaluating copilots

**Testing and evaluation of AI** poses unique challenges compared to traditional software development. First, there is a paradigm shift in software engineering that is best captured by the "Schillace Laws". It is important to recognize that an AI copilot will consist of computer code and models, and that the boundary between them is flexible and may change over time as model capabilities mature. This is the gist of the first Schillace Law that a model improves, but the code doesn't [26]. An implication of this law is that testing code and testing models have important differences. Traditional software testing techniques such as unit tests rely on deterministic output from a system, while the system we are measuring and the tool we use to measure it may be non-deterministic, and thus require appropriate methodologies for proper measurement. Secondly, despite the natural language output, many of the AI systems available in the market are complex and, by nature, a black box to users and copilot developers, who do not yet have a feasible way to understand system output from a causal perspective. There is no guarantee for assertions like "If I input x, system should output y" to work deterministically. Thirdly, the dynamic nature of interactions between the system and the user makes the possible output space very large. The multi-turn conversations for which copilots are typically designed make collecting diverse, representative data for testing and evaluation a difficult task. Lastly, there is a set of biases associated with the use of AI evaluators such as self-bias (preferring response from the same model versus other models or humans), and position bias (preferring certain arbitrary ranking over multiple choices). All these challenges demand that care be taken when using AI evaluators for testing and evaluation of copilots.

**A safety and quality trade-off** is empirically found [27] to exist in training a foundational model. Such a trade-off implies that increasing the safety performance of a model may decrease the model capabilities, or vice versa. This means that some form of the said trade-off has been made for the user of the model. As previously mentioned, the AI copilot speaks for the brand and the value to customers. As a result, before rolling out the copilot, it is important to re-evaluate this trade-off and align the copilot according to the value preference, perhaps using the lifecycle of responsible AI practices mentioned above [4]. In our experience, we recognize that safety and quality evaluations need to be considered jointly for product decisions, along with other factors. When the effectiveness of the AI capabilities is at odds with the set of human values associated with safety, it is sometimes helpful to quantify this trade-off with measurements and plot a graph to find an "optimal" trade-off that aligns with the intended values. For example, if there are multiple versions of a copilot that have been developed, they can be represented by different trade-off points on a Pareto frontier, a graph that measures the best trade-offs among several factors, such as quality, safety, and cost. Quantifying these tradeoffs helps make an informed choice for the best version of the copilot to release in alignment with value preferences. For example, given a certain threshold in quality, the best copilot might be the version with the highest "safety" scores according to the predetermined definitions. Sometimes, after better mitigations are in place, an even better frontier of the quality-safety trade-off may be obtained, showing improved quality performance given the same safety thresholds. It is important to always ground this discussion in the appropriate socio-economic context of the product, acknowledging that sometimes it may not be possible to find the "best" trade-off. The point of this practice, nevertheless, is to recognize and have the choice to reassert the value preference of the product, instead of naively assuming that the foundational model is necessarily ready "straight out of the oven." Overall, this practice provides another opportunity to design human-AI interaction to augment the role of human agency.

## 4  Conclusion

In this paper, we have presented some business and technical considerations for the design and evaluation of two AI copilots in the retail domain. Our systematic approaches to the discussion have been motivated by the business goals of managing desired outcomes and unintended consequences. Drawing on our experience building the retail copilot templates, we have presented our approaches to designing for human-AI collaboration and ensuring the quality and safety of the AI-generated outputs, with the aim of aligning copilot responses with the product values. We have also shared some of the responsible AI practices that we used to test and mitigate the risks of deploying AI-powered copilots in enterprise settings. We hope that our experiences and insights can inspire and inform business leaders and practitioners who are interested in building enterprise-grade copilots.

We believe that the two copilot templates are just two examples of many promising examples of how AI can augment human productivity in various tasks and scenarios. However, we also acknowledge that there are still many open questions and challenges that need to be addressed in order to ensure the ethical and trustworthy use of AI copilots in real-world contexts. We hope that by sharing our work and learnings, we can motivate and advocate for the continued investment in systematic, human-centric approaches to copilot design and evaluation.

## AI disclosure note

In the process of preparing this paper the authors used the following AI models and tools for literature research, structuring of the content and editing: copilot for Microsoft 365 (https://copilot.microsoft.com/), ChatGPT 3.5 and ChatGPT 4 (via https://chat.openai.com/), Claude.ai (via https://claude.ai/chats), Claude 3 models (https://console.anthropic.com/workbench/ API – with the following models claude-3-haiku-20240307, claude-3-sonnet-20240229, claude-3-opus-20240229), Gemini (https://gemini.google.com/app), Perplexity.ai (https://www.perplexity.ai/). The authors attest that they hold full responsibility for the text and factual correctness in the paper above, and have fully reviewed references and edited text outputs from AI systems.